\documentclass[epj,twocolumn]{webofc}
\usepackage[varg]{txfonts}   
\woctitle{Hadron Collider Physics symposium 2012}

\usepackage{color}
\usepackage{xspace}
\usepackage{amsmath}
\usepackage{ifthen} 
\newboolean{uprightparticles}
\setboolean{uprightparticles}{false} 

\def\ux85 {\mbox{UX85}\xspace}

\ifthenelse{\boolean{uprightparticles}}%
{

 \def\Pmu         {\ensuremath{\upmu}\xspace}

 \def\Ptau        {\ensuremath{\uptau^{-}}\xspace}

 \def\PDelta      {\ensuremath{\Delta}\xspace}                 
 \def\PXi      {\ensuremath{\Xi}\xspace}                 
 \def\PLambda      {\ensuremath{\Lambda}\xspace}                 
 \def\PSigma      {\ensuremath{\Sigma}\xspace}                 
 \def\POmega      {\ensuremath{\Omega}\xspace}                 
 \def\PUpsilon      {\ensuremath{\Upsilon}\xspace}                 
 
 \def\PB      {\ensuremath{\mathrm{B}}\xspace}                 
                  
 \def\PD      {\ensuremath{\mathrm{D}}\xspace}

 \def\PK      {\ensuremath{\mathrm{K}}\xspace}

 \def\Pi      {\ensuremath{\mathrm{i}}\xspace}

 \def\Ps      {\ensuremath{\mathrm{s}}\xspace}

}
{

 \def\Pmu         {\ensuremath{\mu}\xspace}

 \def\Ptau        {\ensuremath{\tau^-}\xspace}

 \mathchardef\PDelta="7101
 \mathchardef\PXi="7104
 \mathchardef\PLambda="7103
 \mathchardef\PSigma="7106
 \mathchardef\POmega="710A
 \mathchardef\PUpsilon="7107
                  
 \def\PB      {\ensuremath{B}\xspace}                 
                  
 \def\PD      {\ensuremath{D}\xspace}

 \def\PK      {\ensuremath{K}\xspace}

 \def\Pi      {\ensuremath{i}\xspace}

 \def\Ps      {\ensuremath{s}\xspace}

}

\def\mup        {\ensuremath{\Pmu^+}\xspace}
\def\mun        {\ensuremath{\Pmu^-}\xspace} 

\def\squark    {\ensuremath{\Ps}\xspace}

\def\kaon  {\ensuremath{\PK}\xspace}
  \def\Kbar  {\kern 0.2em\overline{\kern -0.2em \PK}{}\xspace}

\def\Kz    {\ensuremath{\kaon^0}\xspace}
\def\Kzb   {\ensuremath{\Kbar^0}\xspace}
\def\KzKzb {\ensuremath{\Kz \kern -0.16em \Kzb}\xspace}
\def\Kp    {\ensuremath{\kaon^+}\xspace}
\def\Km    {\ensuremath{\kaon^-}\xspace}

\def\KpKm  {\ensuremath{\Kp \kern -0.16em \Km}\xspace}

\def\Kstarz  {\ensuremath{\kaon^{*0}}\xspace}
\def\Kstarzb {\ensuremath{\Kbar^{*0}}\xspace}

  \def\Dbar    {\kern 0.2em\overline{\kern -0.2em \PD}{}\xspace}
\def\D       {\ensuremath{\PD}\xspace}

\def\Dz      {\ensuremath{\D^0}\xspace}
\def\Dzb     {\ensuremath{\Dbar^0}\xspace}
\def\DzDzb   {\ensuremath{\Dz {\kern -0.16em \Dzb}}\xspace}
\def\Dp      {\ensuremath{\D^+}\xspace}
\def\Dm      {\ensuremath{\D^-}\xspace}

\def\DpDm    {\ensuremath{\Dp {\kern -0.16em \Dm}}\xspace}

\def\Dstarpm {\ensuremath{\D^{*\pm}}\xspace}

\def\Ds      {\ensuremath{\D^-_\squark}\xspace}

\def\B       {\ensuremath{\PB}\xspace}
  \def\Bbar    {\kern 0.18em\overline{\kern -0.18em \PB}{}\xspace}

\def\Bu      {\ensuremath{\B^+}\xspace}

\def\Bd      {\ensuremath{\B^0}\xspace}
\def\Bs      {\ensuremath{\B^0_\squark}\xspace}

\def\Bdb     {\ensuremath{\Bbar^0}\xspace}

  \def\Y#1S{\ensuremath{\PUpsilon{(#1S)}}\xspace}

\def\Lbar {\ensuremath{\kern 0.1em\overline{\kern -0.1em\PLambda}}\xspace}

\def\BF         {{\ensuremath{\cal B}\xspace}}

\newcommand{\decay}[2]{\ensuremath{#1\!\to #2}\xspace}         
\def\ra                 {\ensuremath{\rightarrow}\xspace}
\def\to                 {\ensuremath{\rightarrow}\xspace}

\def\BdToKstmm    {\decay{\Bd}{\Kstarz\mup\mun}}
\def\BdbToKstmm   {\decay{\Bdb}{\Kstarzb\mup\mun}}

\def\AT#1     {\ensuremath{A_{\mathrm{T}}^{#1}}\xspace}           

\def\Bsmm     {\decay{\Bs}{\mup\mun}}
\def\Bdmm     {\decay{\Bd}{\mup\mun}}

\def\C#1      {\ensuremath{\mathcal{C}_{#1}}\xspace}                       
\def\Cp#1     {\ensuremath{\mathcal{C}_{#1}^{'}}\xspace}                    
\def\Ceff#1   {\ensuremath{\mathcal{C}_{#1}^{\mathrm{(eff)}}}\xspace}        
\def\Cpeff#1  {\ensuremath{\mathcal{C}_{#1}^{'\mathrm{(eff)}}}\xspace}       
\def\Ope#1    {\ensuremath{\mathcal{O}_{#1}}\xspace}                       
\def\Opep#1   {\ensuremath{\mathcal{O}_{#1}^{'}}\xspace}                    

\newcommand{\tev}{\ensuremath{\mathrm{\,Te\kern -0.1em V}}\xspace}
\newcommand{\gev}{\ensuremath{\mathrm{\,Ge\kern -0.1em V}}\xspace}
\newcommand{\mev}{\ensuremath{\mathrm{\,Me\kern -0.1em V}}\xspace}
\newcommand{\kev}{\ensuremath{\mathrm{\,ke\kern -0.1em V}}\xspace}
\newcommand{\ev}{\ensuremath{\mathrm{\,e\kern -0.1em V}}\xspace}
\newcommand{\gevc}{\ensuremath{{\mathrm{\,Ge\kern -0.1em V\!/}c}}\xspace}
\newcommand{\mevc}{\ensuremath{{\mathrm{\,Me\kern -0.1em V\!/}c}}\xspace}
\newcommand{\gevcc}{\ensuremath{{\mathrm{\,Ge\kern -0.1em V\!/}c^2}}\xspace}
\newcommand{\gevgevcccc}{\ensuremath{{\mathrm{\,Ge\kern -0.1em V^2\!/}c^4}}\xspace}
\newcommand{\mevcc}{\ensuremath{{\mathrm{\,Me\kern -0.1em V\!/}c^2}}\xspace}

\def\mub{\ensuremath{\rm \,\Pmu b}\xspace}

\def\invfb   {\ensuremath{\mbox{\,fb}^{-1}}\xspace}

\def\gsim{{~\raise.15em\hbox{$>$}\kern-.85em
          \lower.35em\hbox{$\sim$}~}\xspace}
\def\lsim{{~\raise.15em\hbox{$<$}\kern-.85em
          \lower.35em\hbox{$\sim$}~}\xspace}

\def\tell1  {TELL1\xspace}
\def\ukl1   {UKL1\xspace}

\newcommand{\CLs}{\ensuremath{\textrm{CL}_{\textrm{s}}}\xspace}

\newcommand{\Ks}{\ensuremath{K^0_s}\xspace}
\newcommand{\Kl}{\ensuremath{K^0_L}\xspace}
\newcommand{\Ksmm}{\ensuremath{\Ks\to\mu^+\mu^-}\xspace}
\newcommand{\Klmm}{\ensuremath{\Kl\to\mu^+\mu^-}\xspace}
\newcommand{\Kspipi}{\ensuremath{\Ks\to\pi^+\pi^-}\xspace}
\newcommand{\Dmm}{\ensuremath{\Dz\to\mu^+\mu^-}\xspace}

\newcommand{\Bsmumu}{\ensuremath{\Bs\to\mu^+\mu^-}\xspace}
\newcommand{\Bdmumu}{\ensuremath{\Bd\to\mu^+\mu^-}\xspace}

\newcommand{\Bhh}{\ensuremath{B^0_{(s)}\to h^+h'^-}\xspace}

\newcommand{\BuToKmm}{\ensuremath{B^+\to K^{+}\mu^+\mu^-}\xspace}

\newcommand{\DstDpi}{\ensuremath{\Dstarpm\to\Dz(\to\pi^+\pi^-)\pi^\pm}\xspace}

\newcommand{\bpimumu}{\ensuremath{B^{0(+)} \to \pi^{0(+)} \mu^+ \mu^-}\xspace}
\newcommand{\BdPiMuNu}{\ensuremath{\ensuremath{B^0}\to \pi^- \mu^+ \nu_\mu}\xspace}

\newcommand{\tmmm}{\ensuremath{\tau^-\to \mu^+\mu^-\mu^-}\xspace}
\newcommand{\tpmm}{\ensuremath{\tau^-\to p \mu^-\mu^-}\xspace}
\newcommand{\tpbmm}{\ensuremath{\tau^-\to \bar{p} \mu^+\mu^-}\xspace}
\newcommand{\DsPhiPi}{\ensuremath{D_s^-\to \phi (\mu^+\mu^-) \pi^-}\xspace}

\newcommand{\BRof}[1]{\ensuremath{{\cal B}(#1)}\xspace}

\begin{document}
\title{Rare beauty and charm decays at LHCb}
%
%

\author{Johannes Albrecht\inst{1,2}\fnsep\thanks{\email{albrecht@cern.ch}} on behalf of the LHCb collaboration}

\institute{CERN, Geneva, Switzerland \and TU Dortmund, Dortmund, Germany}

\abstract{%
Rare heavy flavor decays are an ideal place to search for the effects
of potential new particles that modify the decay rates or the Lorentz
structure of the decay vertices.
The LHCb experiment, a dedicated heavy flavour experiment at the LHC
at CERN. It has recorded the
worlds largest sample of heavy meson and lepton decays. The status of the rare decay
analyses with 1\invfb of $\sqrt s = 7\tev $ and 1.1\invfb of $\sqrt s =
8\tev $ of $pp$--collisions collected by the LHCb experiment in 2011
and 2012 is reviewed. 
The worlds most precise measurements of the angular structure of
\BdToKstmm and \BuToKmm decays is discussed, as well as the isospin
asymmetry measurement in $\decay{B}{\kaon^{(*)} \mup\mun}$ decays. The
first evidence for the very rare decay \Bsmm is presented together
with the most stringent upper exclusion limits on the branching
fraction of decays of \Bd, \Dz and \Ks mesons into two muons. 
This note finishes with the discussion of searches for lepton number
and lepton flavor violating $\tau$ decays.  

}
\maketitle

\section{Introduction}
\label{sec:intro}

Flavor changing neutral current (FCNC) processes are forbidden at tree
level in the Standard Model (SM), but can proceed via loop level
electroweak penguin or box diagrams. In extensions to
the SM, new virtual particles can enter in these loop level diagrams,
modifying the decay rate or Lorentz structure of the
decay vertex. Possible deviations from the SM predictions of these
observables could lead to the discovery of yet unknown phenomena.
The search for these deviations is a complementary approach to direct searches at general purpose 
detectors and can give sensitivity to new particles at higher 
mass scales than those accessible directly. 

This article reviews some of the most sensitive probes for possible
extensions of the Standard Model that were measured by the LHCb
collaboration at the time of the HCP conference (November 2012). 
Most measurements use a dataset of 1\invfb of $\sqrt s
= 7\tev$ of $pp$--collisions collected in 2011. The search for
\Bsmm uses a combined dataset of 1\invfb of $\sqrt s = 7\tev$ and
1.1\invfb of $\sqrt s= 8\tev$, recorded in 2011 and 2012.

The first part of the article discusses rare electroweak penguin
transitions of the type\footnote{In this proceedings, the inclusion of
  charge conjugate states are implicit, unless otherwise stated.} 
$b \rightarrow s \mu^{+} \mu^{-}$, which allow stringent  
tests of the Lorentz structure of the electroweak penguin processes.
The second part discusses searches for purely leptonic decays of \Ks,
\Dz and \B mesons, which are particularly sensitive to new scalar
interactions. The last class of analyses discussed is the search for
lepton and baryon number violating $\tau$ decays. 

The implications of the presented measurements on possible extensions
of the SM, most notably supersymmetric extensions, is discussed in a
separate contribution in these proceedings~\cite{tb}.

\section{Electroweak penguin decays}
\label{ewp}

\subsection{Angular analysis and CP asymmetries in  \BdToKstmm decays}

The decay \BdToKstmm has a branching fraction of
$\BRof\BdToKstmm=(1.05^{+0.16}_{-0.13})\times
10^{-6}$~\cite{pdg12}. It 
 allows the construction of several observables with
small hadronic uncertainties, that are sensitive to physics beyond the
Standard Model (see \cite{Ali1991505,Altmannshofer:2008dz} and
references therein).  
\begin{figure*}[!htb]
\centering
\includegraphics[width=0.94\textwidth]{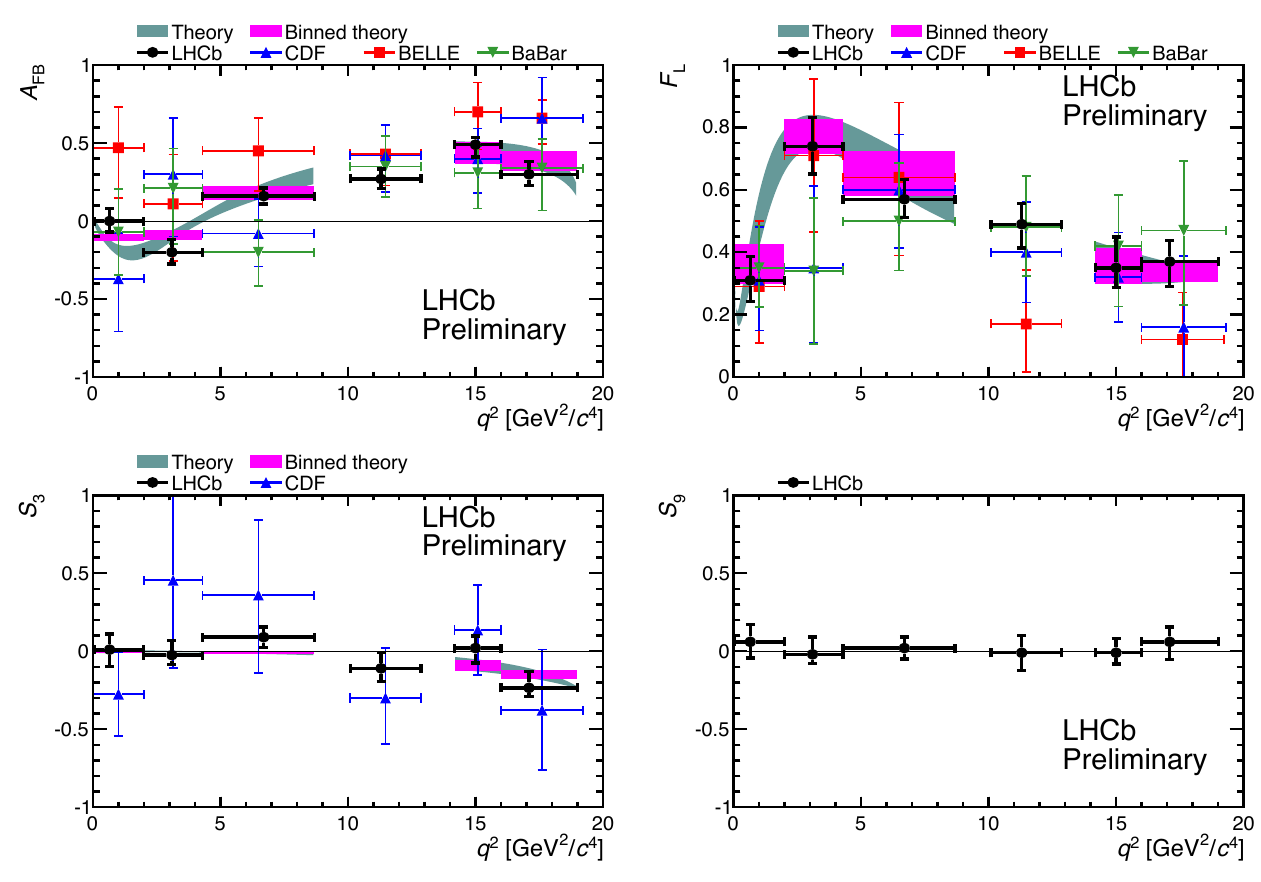}
\caption
{The observables $A_{FB}$, $F_L$, $S_3$ and $S_9$ measured in
  \BdToKstmm decays by the BABAR~\cite{n25}, 
  Belle~\cite{Wei:2009zv}, CDF~\cite{n28} and LHCb~\cite{LHCb-CONF-2012-008}
  experiments. The SM prediction, from Ref.~\cite{Bobeth:2010wg}, is also
  shown. Figure reproduced from Ref.~\cite{LHCb-CONF-2012-008}.}
\label{fig:kst_obs}
\end{figure*} 
The LHCb collaboration performs an angular analysis in bins of 
the squared dimuon invariant mass  
($q^2$) and the three angles $\theta_l$, $\theta_k$ and
$\phi$~\cite{LHCb-CONF-2012-008}. $\theta_l$ is defined as the angle between the \mup and the
\Bd in the dimuon rest frame, $\theta_k$ as angle between the kaon and
the \Bd in \Kstarz rest frame and $\phi$ as angle between the plane
spanned by the dimuon system and the \Kstarz decay plane.

The differential branching ratio as a function of $q^2$ as well as the
following observables 
have been measured (the observables are fully defined in
~\cite{Altmannshofer:2008dz,Kruger:2005ep,Bobeth:2008ij,Bobeth:2010wg}): 
$A_{FB}$, the forward-backward asymmetry of the dimuon system; $F_L$,
the fraction of \Kstarz longitudinal polarization; $S_3$, the
transverse asymmetry, which is also often referred to as
$\frac{1}{2}(1-F_L)A_T^2$ and $S_9$, a $CP$ averaged quantity
corresponding to the imaginary component of the product of the
longitudinal and transverse amplitudes of the \Kstarz.
The measurement of these observables is shown in
Fig.~\ref{fig:kst_obs}, together with the SM prediction and the
previous measurements of other
collaborations~\cite{n25,Wei:2009zv,n28,LHCb-CONF-2012-008}. 
All observables are found to be consistent with each other and with the
SM predictions. The LHCb results are the most 
precise measurements of these observables. 


A particularly sensitive probe for new phenomena is $q_0^2$, the
zero-crossing point of $A_{FB}$. It  is theoretically very clean as
the form factor uncertainties cancel at first order.
The LHCb collaboration has reported the worlds first measurement as
$q_0^2=4.9^{+1.1}_{-1.3}$\,GeV$^2$/c$^4$, in good agreement with the SM
prediction. This measurement strongly disfavours scenarios with a
flipped sign of the Wilson coefficient $C_7$.

\vskip 2mm

The direct CP asymmetry in the \BdToKstmm system,
\begin{equation}
{\cal A_{CP}}
=\frac{\Gamma(\BdbToKstmm)-\Gamma(\BdToKstmm)}{\Gamma(\BdbToKstmm)+\Gamma(\BdToKstmm)}\,
,
\end{equation}
is predicted to be of order $10^{-3}$ in the Standard Model. 
It was measured 1.0\invfb of $7\tev$ data~\cite{LHCb:2012kz} to be 
\begin{equation}
{\cal   A_{CP}}=-0.072\pm0.040_{stat}\pm0.005_{syst}\, ,
\end{equation}
integrated over the six $q^2$ bins. This measurement is consistent
with the Standard Model prediction, it is the most precise measurement
of $\cal A_{CP}$  in \BdToKstmm decays to date.

\subsection{Angular analysis of \BuToKmm decays}

The angular analysis of \BuToKmm decays is performed analogously to
the analysis of \BdToKstmm decays. The angular distribution of
\BuToKmm decays is given as~\cite{Ali:1999mm,Bobeth:2007dw}
\begin{equation}
\frac{1}{\Gamma}\frac{d\Gamma}{d\cos
  \theta_l}=\frac{3}{4}(1-F_H)(1-\cos^2\theta_l)+\frac{1}{2}F_H+A_{FB}\cos\theta_l\, ,
\end{equation}
where $A_{FB}$ denotes the forward backward asymmetry and $F_H$ the
so called flat parameter. The SM predictions for both parameters are very
small. Both $A_{FB}$ and $F_H$ can be significantly enhanced in models
with large operators $C_S^{(\prime)}$ or $C_P^{(\prime)}$.

\begin{figure}[t]
\centering
\includegraphics[width=0.47\textwidth]{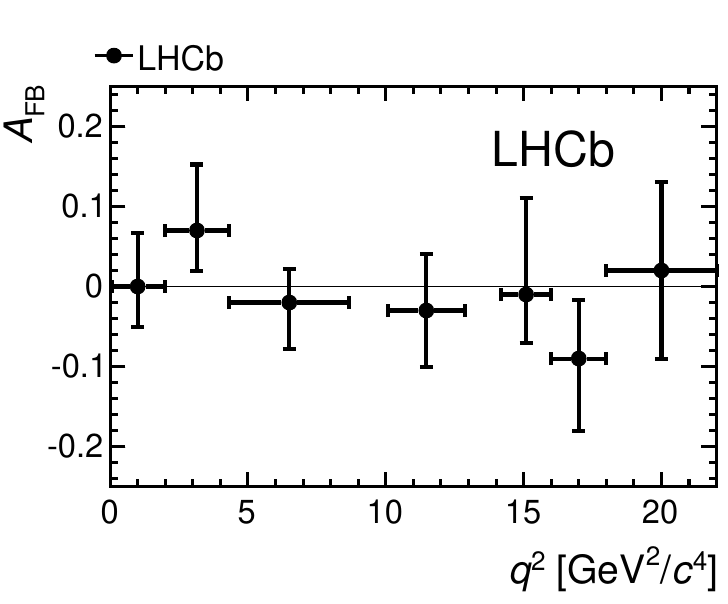}
\includegraphics[width=0.47\textwidth]{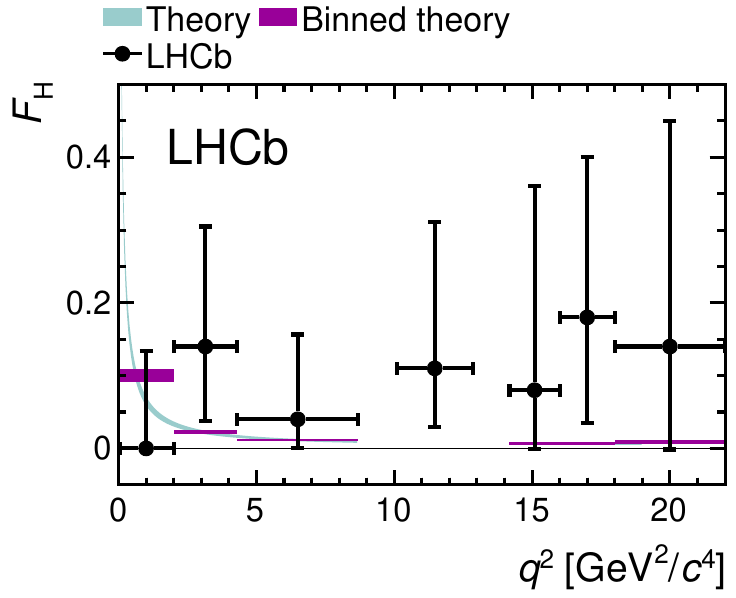}
\caption
{Dimuon forward-backward asymmetry, $A_{FB}$, and the parameter $F_H$
  for \BuToKmm as a function of the dimuon invariant mass squared,
  $q^2$. The SM theory prediction~\cite{Bobeth:2011nj} is also shown.}
\label{fig:kmm}
\end{figure}

The LHCb collaboration has measured $A_{FB}$ and $F_H$ with 1\invfb of
data collected at $\sqrt s=7\tev$~\cite{Aaij:2012vr}, as shown in in
Fig.~\ref{fig:kmm}. The measurement is found in good agreement with
the SM predictions.

\subsection{Isospin asymmetry in $\decay{B}{\kaon^{(*)} \mup\mun}$}

The isospin asymmetry of the decays $\decay{B}{\kaon^{(*)} \mup\mun}$,
$A_I$, is defined as
\begin{equation}
A_I=\frac
{\BRof{\decay{\Bd}{\kaon^{(*)0} \mup\mun}} - \frac{\tau_0}{\tau_+}
  \BRof{\decay{\Bu}{\kaon^{(*)+} \mup\mun}}}
{\BRof{\decay{\Bd}{\kaon^{(*)0} \mup\mun}} + \frac{\tau_0}{\tau_+}
  \BRof{\decay{\Bu}{\kaon^{(*)+} \mup\mun}}}\, ,
\end{equation}
where $\tau_{0,+}$ is the lifetime of the \Bd and \Bu meson
respectively. 
For the $\decay{B}{\kaon^{*} \mup\mun}$ system, in the SM, $A_I$ is
predicted to be $-0.01$~\cite{Feldmann:2002iw}
with 
a slight increase at low values of $q^2$. 
For the $\decay{B}{\kaon \mup\mun}$ system, the SM calculation of
$A_I$ predicts a similar expectation close to
zero~\cite{Khodjamirian:2012rm}. 
The most precise measurement of $A_I$ is performed by the LHCb
collaboration~\cite{Aaij:2012cq}. 
The measurement of $\decay{B}{\kaon^{*} \mup\mun}$ is consistent  with
the SM prediction. The $\decay{B}{\kaon \mup\mun}$ measurement
shows a deviation from zero with a significance of greater than four
standard deviations~\cite{Aaij:2012cq}.


\section{Searches for very rare and forbidden decays}
\label{vrd}

\subsection{Searches in leptonic meson decays}

Decays of \Kz, \Dz and \Bs or \Bd mesons into a muon pair are
discussed in this section. 
The leptonic final state allows
precise calculations of the expected rates and the two muon final
state is experimentally a very clean signature. Both features together
are making these decays very powerful tests of the Standard Model.

\subsubsection{Search for \Ksmm}

The rare decay \Ksmm can give insight into the short-distance
structure of $\Delta S=1$ FCNC transitions. This decay is highly
suppressed in the SM, the predicted branching fraction is
$\BRof{\Ksmm}=(5.0 \pm 1.5) \times
10^{-12}$~\cite{Ecker:1991ru,Isidori:2003ts}. Contributions from possible extensions of
the SM, e.g. from new light scalar particles, can enhance the branching fraction.

The LHCb dataset of 1\invfb contains about $10^{13}$ \Ks decays inside
the detector acceptance. Signal candidates are separated from the
background using BDT based selection. Main sources of residual
background originate from semileptonic decays and \Kspipi decays,
where both pions are misidentified as muons. The latter can be
separated from signal candidates exploiting the excellent mass
resolution of the LHCb spectrometer. 
The contribution of \Klmm is found to be negligible for this analysis.  

The number of expected signal events is evaluated using a relative
normalization to \Kspipi decays. This normalization reduces the
systematic uncertainties which need to be considered in this
analysis. The modified frequentist
method~\cite{Junk:1999kv,0954-3899-28-10-313}, \CLs, is used to
evaluate the consistency of the observed pattern of events with
the background and signal plus background hypotheses. 
The expected upper exclusion limit is at 95\% CL $\BRof{\Ksmm}<1.1
\times 10^{-8}$ and the observed limit is found to be $\BRof{\Ksmm}<1.1
\times 10^{-8}$. This limit constitutes an improvement of a factor 30
with respect to the previous best limit.

\subsubsection{Search for \Dmm}

The \Dmm decay is predicted to be very rare in the Standard
Model~\cite{RareD:Burdman1}: $1 \times 10^{-13} < \BRof \Dmm < 6 \times
10^{-11}$. 
This prediction can be significantly enhanced in MSSM scenarios with R
parity violation, which predicts $\BRof \Dmm \sim 1 \times 10^{-9}$
mediated by a tree level transition~\cite{RareD:Golowich1}.

The LHCb collaboration has performed an analysis using $0.9\invfb$ of
data at $\sqrt s = 7\tev$~\cite{LHCb-CONF-2012-005}.
The background is reduced using a multivariate discriminant based on
geometrical and kinematic information. The signal events are
normalized to the \DstDpi channel, which allows to reduce common
systematic uncertainties.  
The event yield is determined from a two dimensional fit on the dimuon
invariant mass and the difference between the \Dstarpm mass and \Dz
mass.
The observed pattern of events is compatible with the background
expectations and an upper limit on the branching fraction of
$\BRof\Dmm <1.3 \times 10^{-8}$ is determined at 95\% CL, using the $CL_s$ method. 
This is the worlds most stringent limit on this decay.


\subsubsection{Evidence for \Bsmm}

\begin{figure}[htb]
  \begin{center}
    \includegraphics*[width=0.47\textwidth]{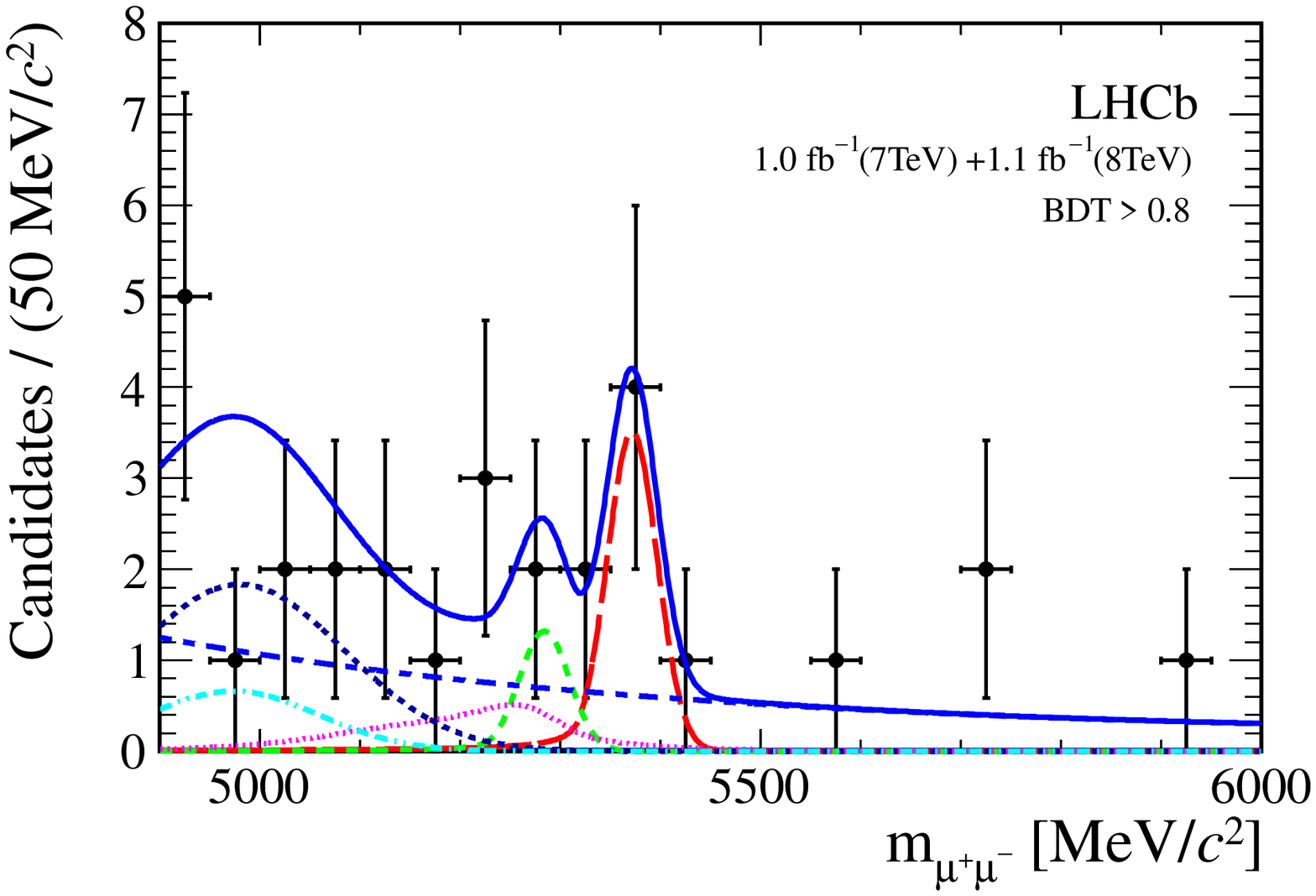}
  \end{center}
\vspace{-8mm}
\caption{Invariant mass distribution of selected \Bsmumu candidates (black points) 
with a high signal likelihood.
The result of the fit is overlaid (blue line) and the different components detailed:
\Bsmumu  (red) and \Bdmumu (violet) SM signals, \Bhh (green), \BdPiMuNu (black) and \bpimumu 
(light blue) exclusive backgrounds.}
\label{fig:fondo_bsd}
\end{figure}

The search for the loop- and helicity suppressed decays \Bsmm and
\Bdmm constitute a very stringent test of possible extensions of the
SM, specially those with an extended scalar sector. The LHCb
collaboration observes an excess of signal candidates in the channel
\Bsmm~\cite{PhysRevLett.110.021801}, which is inconsistent with the
background hypothesis with a significance of 3.5 standard
deviations. This measurement provides the first evidence for this
decay, the measured branching fraction is consistent with the SM expectation. 
The \Bsmm candidates with a high signal likelihood are shown in
Fig.~\ref{fig:fondo_bsd}. The measurement is discussed in more detail in a
separate contribution of these proceedings~\cite{Albrecht:2013wzd}.

\subsection{Search for forbidden \Ptau decays}
Lepton flavor violating (LFV) $\tau^-$ decays only occur in the
Standard Model from neutrino mixing. 
Many extensions beyond the SM predict enhancements, up to observable
values which are close to the current experimental bounds.

\subsubsection{\tmmm}

The neutrinoless decay $\tau^-\to\mu^+\mu^-\mu^-$ is a particular
sensitive mode in which to search for LFV at LHCb as the experimental
signature with the three muon final state is very clean and the
inclusive $\tau^-$ production cross-section at LHCb is very large, 
about $80\mub$. The composition of the $\tau^-$ production can be
calculated from the $b\bar{b}$ and $c\bar{c}$~\cite{Aaij:2011jh} cross-sections measured
at the LHCb experiment and the inclusive branching ratios $b\ra \tau$
and $c\ra \tau$~\cite{pdg10}. About 80\% of the produced
$\tau^-$--leptons originate from $\Ds$ decays. 

LHCb has performed a search for the decay $\tau^-\to\mu^+\mu^-\mu^-$
using 1.0\invfb of data~\cite{LHCb-CONF-2012-015}. 
The signal events are normalized to the \DsPhiPi channel.
The upper limit on the branching fraction was found to be
\begin{equation}
\BF(\tau^-\to\mu^+\mu^-\mu^-)<6.3 \times 10^{-8}
\end{equation}
at 90\,\% C.L, determined using the $CL_s$ method. This has to be
compared with the current best 
experimental upper limit from the Belle collaboration:
$\BF(\tau^-\to\mu^+\mu^-\mu^-)<2.1 \times 10^{-8}$ at 90\,\% C.L.
The large integrated luminosity that will be collected by the upgraded
LHCb experiment will provide a sensitivity corresponding to an upper
limit of a few times $10^{-9}$~\cite{Bediaga:2012py}.

\subsubsection{\tpmm and \tpbmm}

The large $\tau^-$ sample can be exploited by searching for the
baryon number and lepton number violating decays \tpmm and \tpbmm.
Both decays have $\lvert B-L \rvert = 0$ which is predicted by many NP
models. The analysis for these channels~\cite{LHCb-CONF-2012-027}
follows closely that of the \tmmm mode as described above.  

The observed pattern of events for the two decays \tpmm and \tpbmm is
consistent with the background expectation and upper limits on the
branching fraction of 
\begin{eqnarray}
\BRof\tpmm &<& 4.6 \times 10^{-7}\, {\rm ~~and } \\
\BRof\tpbmm &<& 3.4 \times 10^{-7} 
\end{eqnarray}
are obtained, using the $CL_s$ method. These are the first searches
performed for these decays. 
\section{Conclusion}
\label{}

Most scenarios of physics beyond the Standard Model of particle
physics predict measurable effects in the flavor sector, in particular
in rare meson or lepton decays. 
No sign of physics beyond the Standard Model has yet been observed and
stringent limits on its scale have been set.

An angular analysis of the rare electroweak penguin decays \BdToKstmm
and \BuToKmm has been performed as well as a measurement of the
isospin asymmetry in $\decay{B}{\kaon^{(*)} \mup\mun}$ decays. The
measurements are of unprecedented precision and, besides the isospin
asymmetry in agreement with the SM prediction.

Sensitive probes for NP are the purely leptonic decays of \B, \Dz and
\Ks mesons, all of which have been analysed by the LHCb
collaboration. The first evidence on the decay \Bsmm has been measured
and the most stringent upper exclusion limits on the other decays have
been obtained. The LHCb collaboration has also pioneered the analysis
of lepton flavour violating $\tau^-$ decays by performing the first of
such searches at a hadron collider. 

Most measurements presented in this proceedings use 1\invfb of data
collected at $\sqrt s=7\tev$, about one third of the total dataset
recorded by the LHCb experiment. Updates of the analyses with
significantly improved sensitivity are expected in the coming year and
beyond.
\bibliography{article}

\begin{thebibliography}{33}

\bibitem{tb}
T.~Blake, these proceedings  (2013)

\bibitem{pdg12}
{J. Beringer et al.} (Particle Data Group), Phys. Rev. \textbf{D86}, 010001
  (2012)

\bibitem{Ali1991505}
A.~Ali, T.~Mannel, T.~Morozumi, Physics Letters B \textbf{273}, 505  (1991)

\bibitem{Altmannshofer:2008dz}
W.~Altmannshofer, P.~Ball, A.~Bharucha, A.J. Buras, D.M. Straub et~al., JHEP
  \textbf{0901}, 019 (2009), \texttt{0811.1214}

\bibitem{n25}
S.~Akar (BaBar collaboration), Talk at Lake Louise Winter institute  (2012)

\bibitem{Wei:2009zv}
{Wei, J.-T. et al.} (BELLE Collaboration), Phys.Rev.Lett. \textbf{103}, 171801
  (2009), \texttt{0904.0770}

\bibitem{n28}
{T. Aaltonen et al.} (CDF Collaboration), "Phys. Rev. Lett. 108 081807"  (2011)

\bibitem{LHCb-CONF-2012-008}
{LHCb collaboration}, {LHCb-CONF-2012-008}  ({2012})

\bibitem{Bobeth:2010wg}
C.~Bobeth, G.~Hiller, D.~van Dyk, JHEP \textbf{1007}, 098 (2010),
  \texttt{1006.5013}

\bibitem{Kruger:2005ep}
F.~Kruger, J.~Matias, Phys.Rev. \textbf{D71}, 094009 (2005), 21 pages, 16
  figures. Minor typo in Eq. (4.8) corrected: version to appear in Phys. Rev. D
  Report-no: UAB-FT 560, \texttt{hep-ph/0502060}

\bibitem{Bobeth:2008ij}
C.~Bobeth, G.~Hiller, G.~Piranishvili, JHEP \textbf{0807}, 106 (2008),
  \texttt{0805.2525}

\bibitem{LHCb:2012kz}
R.~Aaij et~al. (LHCb Collaboration) (2012), \texttt{1210.4492}

\bibitem{Ali:1999mm}
A.~Ali, P.~Ball, L.~Handoko, G.~Hiller, Phys.Rev. \textbf{D61}, 074024 (2000),
  \texttt{hep-ph/9910221}

\bibitem{Bobeth:2007dw}
C.~Bobeth, G.~Hiller, G.~Piranishvili, JHEP \textbf{0712}, 040 (2007),
  \texttt{0709.4174}

\bibitem{Aaij:2012vr}
R.~Aaij et~al. (LHCb Collaboration) (2012), \texttt{1209.4284}

\bibitem{Bobeth:2011nj}
C.~Bobeth, G.~Hiller, D.~van Dyk, C.~Wacker, JHEP \textbf{1201}, 107 (2012),
  \texttt{1111.2558}

\bibitem{Feldmann:2002iw}
T.~Feldmann, J.~Matias, JHEP \textbf{0301}, 074 (2003), \texttt{hep-ph/0212158}

\bibitem{Khodjamirian:2012rm}
A.~Khodjamirian, T.~Mannel, Y.M. Wang (2012), \texttt{1211.0234}

\bibitem{Aaij:2012cq}
R.~Aaij et~al. (LHCb Collaboration), JHEP \textbf{1207}, 133 (2012),
  \texttt{1205.3422}

\bibitem{Ecker:1991ru}
G.~Ecker, A.~Pich, Nucl.Phys. \textbf{B366}, 189 (1991)

\bibitem{Isidori:2003ts}
G.~Isidori, R.~Unterdorfer, JHEP \textbf{0401}, 009 (2004),
  \texttt{hep-ph/0311084}

\bibitem{Junk:1999kv}
T.~Junk, Nucl.Instrum.Meth. \textbf{A434}, 435 (1999), \texttt{hep-ex/9902006}

\bibitem{0954-3899-28-10-313}
A.L. Read, Journal of Physics G: Nuclear and Particle Physics \textbf{28}, 2693
  (2002)

\bibitem{RareD:Burdman1}
G.~Burdman et~al., Phys.Rev. \textbf{D66}, 014009 (2002),
  \texttt{hep-ph/0112235}

\bibitem{RareD:Golowich1}
E.~Golowich et~al., Phys.Rev. \textbf{D79}, 114030 (2009),
  \texttt{arXiv:0903.2830}

\bibitem{LHCb-CONF-2012-005}
{LHCb collaboration}, LHCb-CONF-2012-005  (2012)

\bibitem{PhysRevLett.110.021801}
R.~Aaij et~al. (LHCb Collaboration), Phys. Rev. Lett. \textbf{110}, 021801
  (2013)

\bibitem{Albrecht:2013wzd}
J.~Albrecht, these proceedings  (2013), \texttt{1302.1317}

\bibitem{Aaij:2011jh}
R.~Aaij et~al. (LHCb Collaboration), Eur.Phys.J. \textbf{C71}, 1645 (2011),
  \texttt{1103.0423}

\bibitem{pdg10}
{K. Nakamura et al.} (Particle Data Group), J. Phys. G \textbf{37}, 075021
  (2010)

\bibitem{LHCb-CONF-2012-015}
{LHCb collaboration}, {LHCb-CONF-2012-015}  ({2012})

\bibitem{Bediaga:2012py}
I.~Bediaga et~al. (LHCb collaboration) (2012), \texttt{1208.3355}

\bibitem{LHCb-CONF-2012-027}
{LHCb collaboration}, LHCb-CONF-2012-027  (2012)

\end{thebibliography}


\begin{thebibliography}{1}

\bibitem{Albrecht:2012hp}
J.~Albrecht, Mod. Phys. Lett. A \textbf{27} (2012), \texttt{1207.4287}

\end{thebibliography}


\end{document}